\documentclass{aastex701}
\usepackage{amsmath}

\begin{document}

\title{Distinct Modes of Pulsar Glitch Activity Revealed by the Waiting-Time--Amplitude Morphology}

\author[orcid=0009-0002-5615-4810,sname=Liu]{Wen-Tong Liu}
\affiliation{State Key Laboratory of Radio Astronomy and Technology, Xinjiang Astronomical Observatory, CAS, 150 Science 1-Street, Urumqi, Xinjiang, 830011, P. R. China}
\affiliation{University of Chinese Academy of Sciences, 19A Yuquan Road, Beijing 100049, P. R. China.}
\email[show]{liuwentong@xao.ac.cn}

\author[orcid=0000-0003-1473-5713,sname=Wang]{Wei-hua Wang}
\affiliation{College of Mathematics and Physics, Wenzhou University, Wenzhou 325035, P. R. China.}
\email[show]{wang-wh@wzu.edu.cn}

\correspondingauthor{Xia Zhou}
\author[orcid=0000-0003-4686-5977,gname=Xia, sname=Zhou]{Xia Zhou} 
\affiliation{State Key Laboratory of Radio Astronomy and Technology, Xinjiang Astronomical Observatory, CAS, 150 Science 1-Street, Urumqi, Xinjiang, 830011, P. R. China}
\affiliation{Xinjiang Key Laboratory of Radio Astrophysics, 150 Science 1-Street, Urumqi 830011, P. R. China.}
\email[show]{zhouxia@xao.ac.cn}

\author[orcid=0000-0002-0069-831X,gname=Fei-fei, sname=Kou]{Fei-fei KOU} 
\affiliation{State Key Laboratory of Radio Astronomy and Technology, Xinjiang Astronomical Observatory, CAS, 150 Science 1-Street, Urumqi, Xinjiang, 830011, P. R. China}
\affiliation{Xinjiang Key Laboratory of Radio Astrophysics, 150 Science 1-Street, Urumqi 830011, P. R. China.}
\email[show]{koufeifei@xao.ac.cn}

\begin{abstract}
Pulsar glitches display a wide variety of behaviors. They are commonly labeled as ``Vela-like'' or ``Crab-like,'' yet these qualitative categories do not provide reproducible quantitative boundaries. 
In this work, we constructed an objective classification scheme based on the joint distribution of backward waiting time $\Delta t_{-}$ and fractional glitch amplitude $\Delta\nu/\nu$. 
A consolidated sample of 349 glitches from 22 pulsars was compiled by cross-matching the Jodrell Bank and ATNF glitch databases. 
For each source, we applied two-dimensional kernel density estimation (KDE) together with highest-density regions (HDRs) to derive geometric descriptors, including the HDR area, anisotropy ratio, and out-of-HDR fraction. 
Hierarchical clustering of the KDE–HDR descriptors revealed two morphology classes independent of visual inspection. In this convention, the Compact class is associated with small HDR areas, elevated anisotropy, and limited peripheral occupancy. The Extended class is associated with broader HDR occupancy and stronger low-density extensions. 
We found that bootstrap resampling and hyperparameter sensitivity tests confirmed this partition for most sources. 
Nevertheless, two prolific pulsars (PSR~J0537$-$6910 and PSR~J1341$-$6220) were found to occupy intermediate positions. 
Logistic regression identified the mean backward waiting time $\langle\Delta t_{-}\rangle$ as the strongest class predictor (in-sample $\mathrm{AUC}=0.88$; leave-one-out $\mathrm{AUC_{CV}}=0.59$, reflecting the limited sample size). 
Additionally, a parallel analysis using $\Delta t_{+}$ recovered the same two-class structure but with reduced stability. 
The data suggest that the Compact class is consistent with near-complete reservoir depletion, whereas the Extended class reflects partial, avalanche-like releases. 
As a result, short-term measurements of $G$ in Extended-class pulsars may underestimate the true crustal superfluid fraction.
\end{abstract}




\section{Introduction}
\label{sec:intro}

Pulsars were first recognized as rapidly rotating, magnetized neutron stars through early radio observations \citep{hewish_observation_1968,gold_rotating_1968}. The way these objects spin down over long timescales offers a direct window into the internal dynamics of neutron stars \citep{haskell_models_2015,Antonopoulou2022}. The physical implications are broad, ranging from nuclear microphysics all the way to gravitational-wave science \citep{nitu_periodicity_2024,zhou_dependence_2017,haskell_glitching_2024,2025arXiv251217990T}.

When pulse times of arrival are measured with high precision, two main forms of rotational irregularity become apparent. One is timing noise---essentially a stochastic red-noise process that sits on top of measurement white noise. The other is glitches. These are sudden spin-up events, and they are usually followed by some degree of relaxation \citep{hobbs_timingnoise_2010,lower_utmost_2020}. Since the first glitches were detected in the Vela and Crab pulsars \citep{reichley_observed_1969,radhakrishnan_detection_1969}, the number of reported events has grown considerably. To date, more than seven hundred glitches have been documented in over two hundred sources \citep{basu_jodrell_2022,manchester_australia_2005,zubieta_timing_2024,liu_multiband_2025,espinoza_j0537_2026}. This rich observational diversity carries information about neutron-star interiors. In particular, how angular momentum is stored, redistributed, and eventually lost can be traced through differences in glitch size, recurrence interval, and post-glitch recovery behavior.

From an observational standpoint, two broad categories are commonly recognized. Vela-like pulsars tend to undergo large glitches ($\Delta\nu/\nu \sim 10^{-7}\text{--}10^{-6}$) at quasi-periodic intervals of several years. The relaxation that follows is slow and generally incomplete \citep{flanagan_rapid_1990,dodson_two_2002,espinoza_small_2021,montoli_bayesian_2020}. Crab-like sources, on the other hand, produce smaller events ($\Delta\nu/\nu \sim 10^{-9}\text{--}10^{-8}$) that occur at irregular intervals and are followed by rapid, multi-stage recovery \citep{espinoza_study_2011,alpar_postglitch_1993,zheng_persistent_2025,huang_crabglitch_2025}. Although these labels do capture meaningful empirical differences, they fail to establish quantitative boundaries. This makes systematic comparison across sources difficult. It also complicates efforts to connect observed glitch behavior with the physics that drives it.

As the available samples have grown, additional structure at the population level has become visible. Glitch-size distributions appear consistent with bimodal or mixture forms. Waiting times, meanwhile, split between Poisson-like behavior in some objects and quasi-periodic patterns in others \citep{espinoza_study_2011,melatos_avalanche_2008,Fuentes2019,arumugam_classification_2023,eya_glitch_2024,zhu_glitches_2025}. To interpret this structure, one needs predictive trigger models---specifically, models that can account for the joint distribution of glitch sizes and waiting times. Two main classes of models have been put forward.

In starquake models, stress gradually accumulates in the solid crust until brittle failure occurs \citep{ruderman_neutron_1969,baym_neutron_1971,alpar_expectancy_1994,rencoret_revisiting_2021,chugunov_breaking_2010}. The elastic energy available is limited, so these models tend to favor smaller events. They also predict a fairly narrow range of amplitudes, with only weak coupling to waiting-time statistics. Superfluid models take a different approach. Here, glitches arise from the collective unpinning of quantized vortices that are pinned within the inner crust~\citep{baym_superfluidity_1969,anderson_pulsar_1975,pines_superfluidity_1985,pines_pinned_1980,alpar_vortex_1984,glampedakis_hydrodynamical_2009,piekarewicz_pulsar_2014,haskell_models_2015,line,marmorini_pulsar_2024,howitt_nonparametric_2018,layek_glitches_2023}. Because the angular-momentum transfer depends on the lag that builds up between the superfluid and the crust before each event, the glitch amplitude becomes linked to the pre-event waiting time. A wider range of sizes and recurrence behaviors is therefore possible. To distinguish between these two mechanisms, observables that jointly encode information on both glitch size and waiting time are needed.

The traditional Vela-like/Crab-like taxonomy, however, remains qualitative. It is also source-specific, and it does not provide a reproducible quantitative criterion that would allow systematic comparison across the pulsar population. Without such a criterion, the physical origin of the observed diversity stays uncertain. It could reflect genuinely distinct physical regimes, or it might instead represent a continuum shaped by varying stellar parameters. What is needed, then, is a data-driven framework that operates in the joint space of glitch size and recurrence time, so that the observational diversity can be translated into meaningful constraints on trigger mechanisms.

In the present work, we take a geometric approach in the backward-waiting-time--fractional-size plane $(\Delta t_{-},\;\Delta\nu/\nu)$. Here $\Delta t_{-}$ denotes the time elapsed since the preceding glitch. We also define the forward waiting time as $\Delta t_{+,i} = t_{i+1} - t_{i}$, that is, the interval from glitch~$i$ to the subsequent event. We chose the backward interval as our primary variable because it conditions each glitch on the pre-event state of the system, thereby connecting the observed amplitude to the angular-momentum reservoir~\citep{melatos_avalanche_2008,Carlin2019MNRAS,howitt_nonparametric_2018}. The forward interval $\Delta t_{+}$, by contrast, is more sensitive to post-glitch recovery dynamics. A comparative analysis using $\Delta t_{+}$ can be found in Appendix~\ref{app:forward_classification}.

For every source with enough recorded glitches, we constructed a two-dimensional kernel density estimate (KDE) on this plane and extracted the highest-density region (HDR) that encloses a fixed probability mass. From the HDR, a compact set of morphological descriptors was then derived. These descriptors quantify the extent, shape, and concentration of the glitch distribution, along with the prominence of low-density excursions outside the core. Because no predefined functional form is assumed, direct comparisons across sources become possible. By applying hierarchical clustering in this descriptor space and combining it with logistic regression against spin-down properties, we identified a robust two-class morphological structure within the pulsar population. We also found that the mean backward waiting time $\langle\Delta t_{-}\rangle$ serves as the strongest single predictor of morphological class. This finding establishes a quantitative link between the pre-glitch accumulation timescale and the overall character of the glitch distribution.

This paper is organized as follows. In Section~\ref{sec:data_methods}, we describe the data selection procedure, the kernel density estimation framework, and the clustering methodology. Section~\ref{sec:results} presents results on per-source morphology, population-level classification, and the statistical associations with physical drivers. We discuss the physical interpretation of the morphological classes, their implications for the superfluid reservoir, and certain limitations in Section~\ref{sec:discussion}. Section~\ref{sec:conclusion} summarizes our main conclusions.

\section{Data and Methodology} 
\label{sec:data_methods}
 
\subsection{Data Selection and Preparation}
\label{sec:data_prep}
 
A consolidated glitch catalog was constructed by cross-matching the Jodrell Bank Observatory (JBO) pulsar glitch catalog\footnote{\url{http://www.jb.man.ac.uk/pulsar/glitches.html}} \citep{basu_jodrell_2022} against the ATNF glitch table\footnote{\url{https://www.atnf.csiro.au/research/pulsar/psrcat/glitchTbl.html}} \citep{manchester_australia_2005}. Pulsar parameters were taken from the ATNF pulsar catalog\footnote{\url{https://www.atnf.csiro.au/research/pulsar/psrcat/}} \citep{manchester_australia_2005}. For several key sources, we also incorporated glitches that were reported recently in the literature, so as to ensure that the waiting-time statistics are as complete as possible~\citep{ho_new_2026}. All glitches reported through March 2026 are included in the dataset. The JBO catalog records, for each published glitch and for regularly updated Lovell telescope detections, the glitch epoch in Modified Julian Date together with the fractional steps in spin frequency and its derivative \citep[e.g.,][]{espinoza_study_2011}. Throughout this work, we adopt the glitch definition given by \citet{espinoza_study_2011}.
 
Two glitch parameters form the basis of our analysis: the relative glitch amplitude ($\Delta\nu/\nu$) and the backward waiting time ($\Delta t_{-}$). We define the backward waiting time as the interval from the immediately preceding glitch to the current one, $\Delta t_{-,i}=t_i-t_{i-1}$ for $i\ge 2$. The first recorded glitch for each pulsar has no predecessor, and such entries are therefore excluded. From a physical standpoint, $\Delta t_{-}$ represents the time that is available for the system to reload between consecutive glitches.
 
\begin{deluxetable*}{lclclc}
\tablecaption{Glitch counts for the 22 pulsars included in the final analysis ($N_g \ge 6$). Data are derived from the JBO glitch catalog \citep{basu_jodrell_2022} and ATNF pulsar catalog \citep{manchester_australia_2005}.
\label{tab:glitch_counts_threecol}}
\tablewidth{0pt}
\tablehead{
\colhead{Pulsar} & \colhead{$N_{\rm g}$} &
\colhead{Pulsar} & \colhead{$N_{\rm g}$} &
\colhead{Pulsar} & \colhead{$N_{\rm g}$}
}
\startdata
\cutinhead{Pulsars with $N_{\rm g}>10$}
PSR J0835$-$4510 (Vela) & 26 &
PSR J0534+2200 (Crab)   & 32 &
PSR J0537$-$6910        & 68 \\
PSR J1801$-$2304        & 15 &
PSR J1341$-$6220        & 35 &
PSR J0631+1036          & 17 \\
PSR J1048$-$5832        & 13 &
PSR J1413$-$6141        & 14 &
PSR J1740$-$3015        & 40 \\
PSR J1825$-$0935        & 14 & & & & \\
\cutinhead{Pulsars with $6\le N_{\rm g}\le10$}
PSR J0205+6449   & 9 &
PSR J0742$-$2822 & 9 &
PSR J1023$-$5746 & 7 \\
PSR J1420$-$6048 & 8 &
PSR J1645$-$0317 & 7 &
PSR J1731$-$4744 & 8 \\
PSR J1737$-$3137 & 10 &
PSR J1801$-$2451 & 9 &
PSR J1814$-$1744 & 8 \\
PSR J1826$-$1334 & 8 &
PSR J1841$-$0524 & 8 &
PSR J2229+6114   & 9 \\
\enddata
\tablecomments{$N_{\rm g}$ is the total number of glitches before data cleaning. Only pulsars with retained $N_{\rm g}\ge 6$ after data cleaning (Section~\ref{sec:data_prep}) are included. The first glitch lacks $\Delta t_{-}$ and is excluded from waiting-time analyses; further event-level removals (Section~\ref{sec:data_prep}) reduce the effective KDE sample sizes listed in Table~\ref{tab:hdr_params}. The high-$N_{\rm g}$ subset contains 274 glitches from 10 pulsars; the lower subset contains 100 glitches from 12 pulsars, yielding a total of 374 glitches from 22 pulsars.}
\end{deluxetable*}
 
\begin{figure*}[ht!]
\plotone{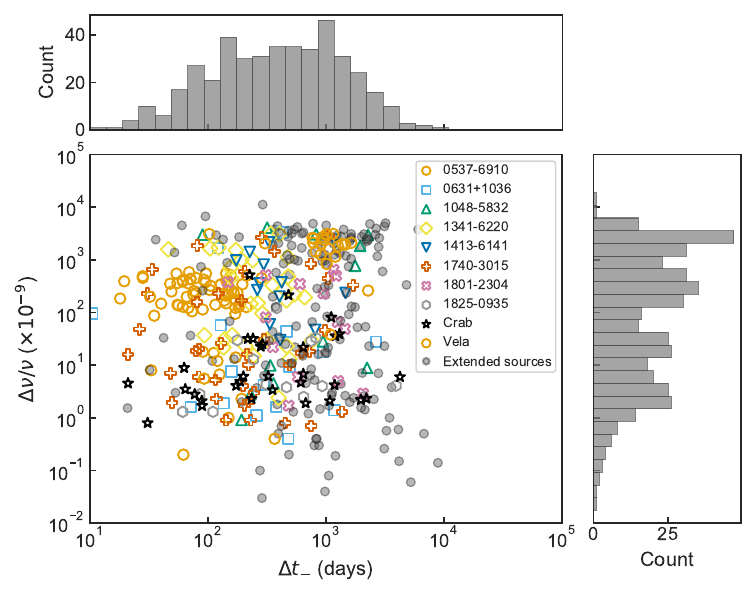}
\caption{Global distribution of the cleaned sample (349 glitches from 22 pulsars) in the backward waiting time versus relative amplitude plane. Each point represents a single glitch. Colored symbols highlight the 10 frequent glitchers, while gray circles denote the remaining sources. The top and right panels show the marginal number distributions of $\Delta t_{-}$ and $\Delta\nu/\nu$, respectively. 
\label{fig:general}}
\end{figure*}

Uniform data cleaning procedures were applied to ensure statistical reliability. We removed duplicate entries, events for which $\Delta\nu/\nu$ was missing, and outliers that could be attributed to extended observational gaps or obvious cataloguing errors. The specific exclusions are listed below; the integer in parentheses refers to the sequential glitch number in the JBO catalog. J2225+6535(1) and J1902+0615(1) were excluded because long gaps inflated the apparent $\Delta t_{-}$. After this filtering step, these two sources were removed from the analysis entirely. J0537$-$6910(1) was also excluded, as it is distorted by incomplete observational coverage. Additionally, J1341$-$6220(2) was dropped because it lacks $\Delta\nu/\nu$. We restricted the morphological analysis to pulsars with $N_{\rm g}\ge 6$ recorded events. This particular threshold was chosen on the basis of a bootstrap stability analysis of the inferred density distributions (Section~\ref{sec:methods_kde}). In the retained sample, values of $\Delta\nu/\nu$ cover three orders of magnitude, from $10^{-9}$ to $10^{-6}$, and $\Delta t_{-}$ ranges from a few days up to $\gtrsim 4\times 10^{3}$ days. Figure~\ref{fig:general} shows how these parameters are distributed, and it illustrates the broad dynamic range of the dataset. Before cleaning, the dataset contained 374 glitches from 22 pulsars (Table~\ref{tab:glitch_counts_threecol}). After cleaning, 349 glitches from the same 22 pulsars remained for analysis (Table~\ref{tab:hdr_params}).
 
\subsection{Kernel Density Estimation}
\label{sec:methods_kde}
 
For individual pulsars, the glitch data in the $(\Delta t_{-},\;\Delta\nu/\nu)$ plane tend to be sparse and irregularly distributed. Parametric approaches to modeling the joint distribution---bivariate Gaussian models or finite-mixture models, for instance---require assumptions about the functional form that are difficult to justify given the small sample sizes involved. Standard binning techniques such as two-dimensional histograms are also problematic. They are sensitive to bin placement, and for sources with only $N_{\mathrm{g}} \sim 6$--$10$ events, they can easily introduce artificial structure into the estimated distribution. For these reasons, we turned to kernel density estimation (KDE). This is a nonparametric method that reconstructs the underlying probability distribution directly from the observed data, without requiring any particular functional form to be specified in advance. By converting discrete glitch events into a smooth, continuous density field, KDE makes it possible to extract geometric descriptors in a consistent way across all sources. These descriptors---including the highest-density region (HDR) area $A_{\mathrm{HDR}}$, the anisotropy ratio AR, and the out-of-HDR fraction $f_{\mathrm{out}}$---capture the topology of glitch occupancy even when the underlying records are heterogeneous.
 
We begin by defining the data matrix for a pulsar with $n$ glitches as $\mathbf{X} = \{\mathbf{x}_i\}_{i=1}^{n}$, where $\mathbf{x}_i = (\Delta t_{-,i},\;(\Delta\nu/\nu)_i)^{T}$. The sample mean vector is $\boldsymbol{\mu} = n^{-1}\sum_{i}\mathbf{x}_i$. From these data, the sample covariance matrix $\mathbf{S}$ was computed. A small ridge term was then added to give $\mathbf{S}_{\mathrm{reg}} = \mathbf{S} + \epsilon\mathbf{I}$, where $\epsilon = 10^{-10}$, and a Cholesky decomposition $\mathbf{S}_{\mathrm{reg}} = \mathbf{L}\mathbf{L}^{T}$ was performed. Each data point was subsequently mapped into whitened coordinates through the transformation $\mathbf{z}_i = \mathbf{L}^{-1}(\mathbf{x}_i - \boldsymbol{\mu})$.
 
The KDE was fitted in the whitened space with a Gaussian kernel,
$K(\mathbf{u})=(2\pi)^{-1}\exp\!\left(-\tfrac{1}{2}\mathbf{u}^{\mathrm{T}}\mathbf{u}\right)$.
The resulting density estimator takes the form
\begin{equation}
\hat{f}(\mathbf{z})=\frac{1}{n h^{2}}\sum_{i=1}^{n}K\!\left(\frac{\mathbf{z}-\mathbf{z}_{i}}{h}\right),
\end{equation}
in which $h$ is a scalar bandwidth. We set $h$ according to the Silverman scaling rule with a small-sample correction: $h=\left[\max(n-1,\,2)\right]^{-1/6}$. It should be noted that the Silverman rule implicitly assumes a unimodal reference distribution. In sources whose glitch populations are genuinely multimodal---Vela being one example (see Section~\ref{sec:disc_physics})---this choice may over-smooth secondary peaks. Nevertheless, the hyperparameter sensitivity tests presented in Section~\ref{sec:population_classes} show that reducing the bandwidth by half ($h_{\mathrm{mult}}=0.5$) does not qualitatively change the two-class classification for most sources, even though boundary objects may shift. Adopting an adaptive-bandwidth KDE that locally adjusts the smoothing to match the data density would be a natural extension for future work. The density field was evaluated on a regular grid in the original $(\Delta t_{-},\,\Delta \nu/\nu)$ plane by applying the whitening transform $\mathbf{z}=\mathbf{L}^{-1}(\mathbf{x}-\boldsymbol{\mu})$ at each grid point $\mathbf{x}$.
 
To describe the main concentration of each pulsar's glitch distribution, we used the highest-density region (HDR). The adopted probability mass was $q=0.68$, which corresponds to the $1\sigma$ enclosed probability for a two-dimensional Gaussian. This value was chosen because it captures the primary core of each source's glitch activity while leaving out peripheral events that may represent rare or extreme excursions. Formally, the HDR is defined as
\begin{equation}
R_{q}=\left\{(x,y)\,\big|\,\hat{f}(x,y)\ge\lambda_{q}\right\},
\end{equation}
where the density threshold $\lambda_{q}$ is determined by the condition
\begin{equation}
\iint_{R_{q}}\hat{f}(x,y)\,\mathrm{d}x\,\mathrm{d}y = q.
\end{equation}
 
Several morphological parameters were then derived from the HDR. First, the HDR area quantifies how large the main density concentration is:
\begin{equation}
A_{\mathrm{HDR}} = N_{\mathrm{cell}}\,\Delta x\,\Delta y,
\end{equation}
where $N_{\mathrm{cell}}$ is the number of grid cells that fall inside the HDR.
 
Second, we defined an out-of-HDR fraction to measure how many events lie outside the core:
\begin{equation}
f_{\mathrm{out}}=\frac{1}{n}\sum_{i=1}^{n}
\mathbf{1}\!\left[\hat{f}(\mathbf{x}_{i}) < \lambda_{0.68}\right],
\end{equation}
where $\mathbf{1}[\cdot]$ is the indicator function. Third, the shape of the glitch distribution was characterized using an anisotropy ratio ($\mathrm{AR}$) derived from the eigenvalues of the sample covariance matrix $\mathbf{S}$. If $\ell_{1}$ and $\ell_{2}$ are the eigenvalues with $\ell_{1}\ge \ell_{2}$, then
\begin{equation}
\mathrm{AR}=\sqrt{\frac{\ell_{1}}{\ell_{2}}}.
\end{equation}
Finally, the mode of the KDE was identified as the grid point at which the density reaches its maximum: $(x_{\mathrm{mode}},\,y_{\mathrm{mode}})=\arg\max_{(x,y)}\hat{f}(x,y)$.
 
Taken together, these HDR-based quantities---$\lambda_{q}$, $A_{\mathrm{HDR}}$, $f_{\mathrm{out}}$, $\mathrm{AR}$, and the mode location---provide a compact yet informative description of the scale, concentration, and geometry of the glitch distribution for each pulsar.
 
\begin{figure*}
\plotone{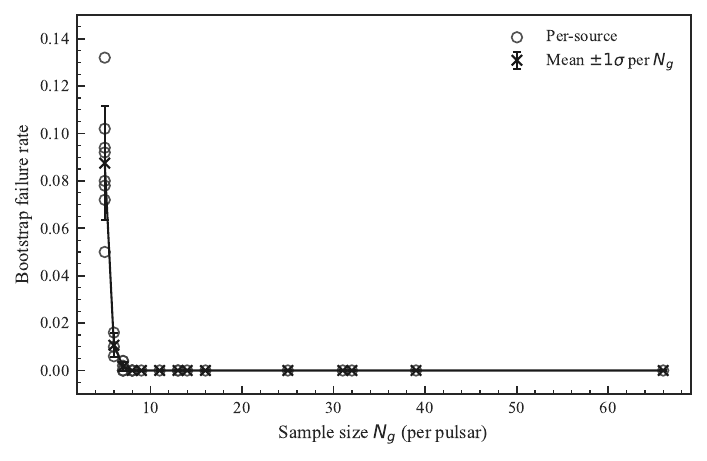}
\caption{Bootstrap failure rate of KDE--HDR reconstruction as a function of per-pulsar sample size $N_{\mathrm{g}}$. Each point represents one source; error bars denote the mean and $\pm1\sigma$ dispersion at fixed $N_{\mathrm{g}}$. The failure rate drops significantly for $N_{\mathrm{g}}\ge 6$, motivating the adoption of this threshold for robust morphological classification.}
\label{fig:failrate}
\end{figure*}
 
We carried out bootstrap resampling for each source to assess the statistical reliability of the HDR-derived parameters. As shown in Figure~\ref{fig:failrate}, the bootstrap failure rate decreases sharply as $N_{\mathrm{g}}$ increases. A clear break is visible between $N_{\mathrm{g}}=5$ and $N_{\mathrm{g}}=6$. Sources with $N_{\mathrm{g}}\ge 6$ consistently produce stable KDE--HDR parameters, and this finding justifies the sample-selection threshold that was introduced in Section~\ref{sec:data_prep}.
 
\subsection{Clustering and Stability Analysis}
\label{sec:methods_cluster}
 
For each pulsar $i$, we constructed a morphological feature vector $\mathbf{x}_{i}=\left(\log_{10}A_{\mathrm{HDR},i},\,\log_{10}\mathrm{AR}_{i},\,f_{\mathrm{out},i}\right)$. All three features were standardized to have zero mean and unit variance before the clustering step.
 
The Hopkins statistic was first computed to check whether the data display a genuine clustering tendency. After confirming this, we applied agglomerative hierarchical clustering using Ward linkage. This method works by successively merging the pair of clusters whose merger causes the smallest increase in total within-cluster variance. The dendrogram was cut at $k=2$, thereby defining two morphological classes.
 
To evaluate the stability of the resulting partition, we took several complementary steps. In each bootstrap realization, the scatter in $\log_{10}A_{\mathrm{HDR}}$, $\log_{10}\mathrm{AR}$, and $f_{\mathrm{out}}$ was recorded, along with the bootstrap failure rate ($f_{\mathrm{fail}}$). Partition quality was assessed through the silhouette score. We further tested whether the classification depends critically on any single feature by repeating the clustering after removing each feature in turn. A bootstrap consensus matrix was also built to assign an object-level stability score to each pulsar. In addition, we repeated the entire feature-extraction and clustering procedure across a grid of hyperparameters, varying both the bandwidth multiplier and the HDR probability level $q$. The agreement between the resulting labeling was measured using the adjusted Rand index (ARI) and the normalized mutual information (NMI).
 
To probe the physical basis of the morphological classes, we fitted a logistic regression model that links the binary class labels to a set of pulsar physical parameters. Denoting the class label of pulsar $i$ by $y_i\in\{0,1\}$ and the vector of standardized physical parameters by $\mathbf{p}_i$, the model is written as
\begin{equation}
P\!\left(y_{i}=1 \mid \mathbf{p}_{i}\right)=\frac{1}{1+\exp\!\left[-\left(\beta_{0}+\boldsymbol{\beta}^{\mathrm{T}}\mathbf{p}_{i}\right)\right]},
\end{equation}
where $\beta_{0}$ is the intercept and $\boldsymbol{\beta}$ is the coefficient vector. In-sample performance was evaluated using the receiver operating characteristic curve and its area under the curve (ROC--AUC). Because the sample is small ($N_{\mathrm{src}}=22$, with six predictors), overfitting is a concern. We therefore also report the leave-one-out cross-validated AUC ($\mathrm{AUC}_{\mathrm{CV}}$) as a more cautious estimate of predictive performance.
 
\section{Results} \label{sec:results}

When applied to the full pulsar sample, the two-dimensional KDE--HDR framework reveals distinct morphological patterns.

\subsection{Per-Source Distribution Characteristics} \label{sec:per_source}

\begin{figure*}[ht!]
\plotone{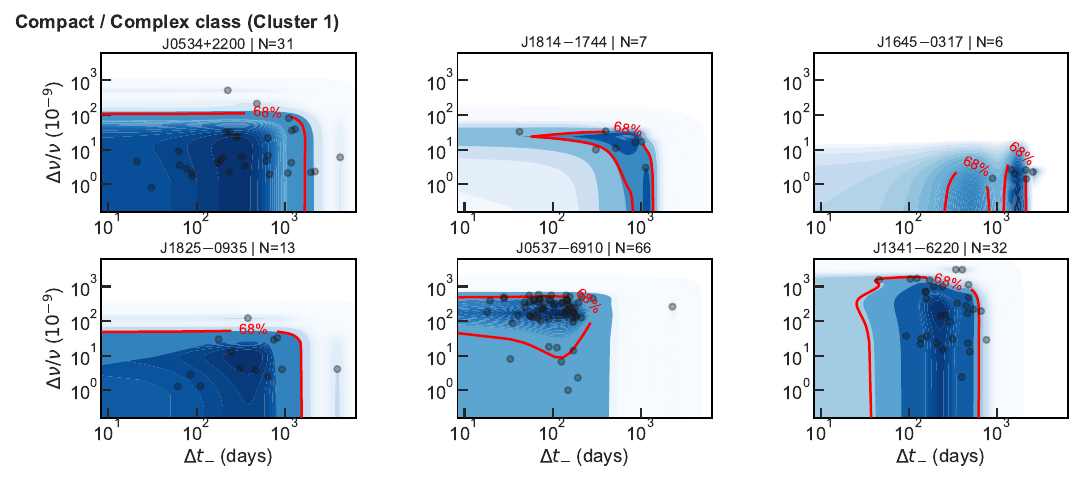}
\caption{KDE--HDR maps for the six pulsars assigned to Cluster~1. Color shading shows the relative KDE density, white circles mark individual
glitches, and red contours enclose the $q=0.68$ HDR. }
\label{fig:kde_c1}
\end{figure*}

\begin{figure*}[ht!]
\plotone{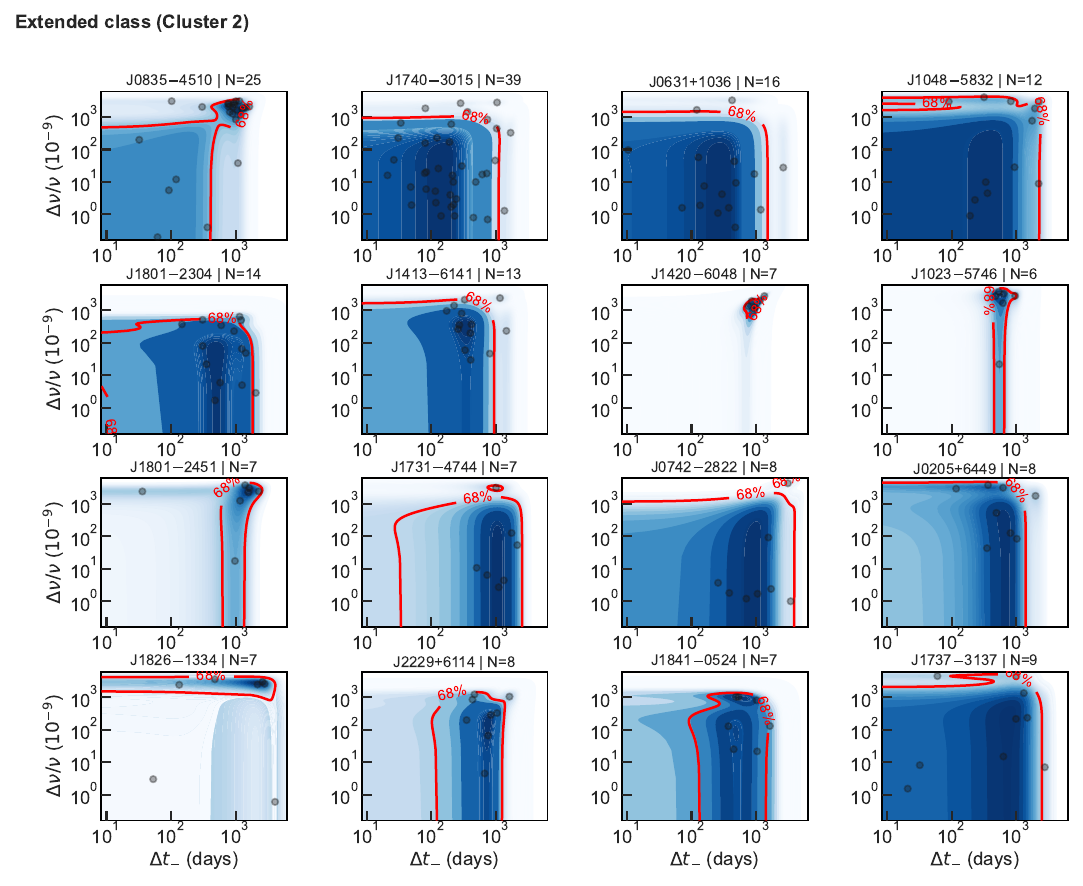}
\caption{KDE--HDR maps for the 16 pulsars assigned to Cluster~2,
the Extended class. Symbols, density shading, and HDR contours follow
Figure~\ref{fig:kde_c1}.}
\label{fig:kde_c2}
\end{figure*}

Figures~\ref{fig:kde_c1} and \ref{fig:kde_c2} provide an overview of the per-source KDE--HDR maps. To make their connection with the population-level analysis clearer, the panels are arranged according to the two hierarchical-clustering branches discussed below rather than by the number of recorded glitches.

The maps exhibit substantial diversity even within a given branch. The class assignments are therefore not based on visual resemblance between
individual panels. They are obtained from Ward clustering of the standardized
descriptor vectors $(\log A_{\rm HDR},\log{\rm AR},f_{\rm out})$. 

The KDE maps can be described in terms of three directly observable features: the extent of the 68\% HDR, the directional elongation of that region, and the presence of events or secondary density concentrations outside the main HDR. These visual features correspond, respectively, to $A_{\rm HDR}$, AR, and $f_{\rm out}$, although the numerical values used in the clustering are calculated from the KDE--HDR analysis rather than assigned by visual inspection.

The Crab pulsar illustrates why these features need to be distinguished. Its 68\% HDR is concentrated at small amplitudes ($\Delta\nu/\nu\lesssim10^{-8}$), but it still spans short-to-moderate backward waiting times ($\Delta t_{-}\sim10^{2}$--$10^{3}$\,d). Its relatively small $A_{\rm HDR}$ therefore arises primarily from confinement in the amplitude direction, rather than from an equally narrow distribution along both axes. Vela instead has its main density concentration near $\Delta t_{-}\sim10^{3}$\,d and $\Delta\nu/\nu\sim10^{-6}$, together with a lower-density extension toward shorter waiting times. PSR~J0537$-$6910 contains several local density maxima and the largest out-of-HDR fraction in the sample, whereas PSR~J1801$-$2304 and PSR~J1048$-$5832 also show secondary density concentrations within broader occupancy fields.

Some sources can appear visually similar while differing in their numerical descriptors. For example, PSR~J1341$-$6220 and PSR~J1413$-$6141 both show a dominant region elongated approximately along the diagonal of the $\Delta t_{-}$--$\Delta\nu/\nu$ plane, yet they are assigned to different clustering branches. Conversely, sources within the same branch need not have identical KDE maps. The population-level classes introduced in Section~\ref{sec:population_classes} consequently reflect the joint standardized descriptor vector $(\log A_{\rm HDR},\log{\rm AR},f_{\rm out})$, rather than any single visual feature.

Before moving on to population-level clustering, we examined whether any individual pulsar contains discrete sub-populations of glitch events. The Hopkins statistic $H$ (Appendix~\ref{app:hopkins}; Table~\ref{tab:hopkins}) was used to test for intrinsic intra-source clustering. For most sources, we obtained $H\approx 0.5$, which is consistent with a single continuous distribution---one that may contain smooth gradients or extended tails, but not distinct sub-clusters. Vela stands out as a notable exception. Its Hopkins statistic of $H=0.66$ provides evidence for genuine bimodal substructure within the glitch population, a finding that deserves further investigation in the context of multi-component superfluid models. On the whole, these diagnostics confirm that global geometric descriptors ($A_{\rm HDR}$, AR, $f_{\rm out}$) offer an appropriate summary of per-source morphology for the great majority of objects. The apparent substructures visible in several of the KDE maps are better understood as smooth density gradients rather than as discrete sub-clusters.

\begin{deluxetable*}{lccccccccc}[htbp]
\tablecaption{KDE--HDR morphological parameters for the analyzed pulsars.\label{tab:hdr_params}}
\tablewidth{0pt}
\tablehead{
\colhead{Source} &
\colhead{$N_{\rm g}$} &
\colhead{$h$} &
\colhead{$\lambda_{q}$} &
\colhead{$A_{\rm HDR}$} &
\colhead{$f_{\rm out}$} &
\colhead{$x_{\rm mode}$} &
\colhead{$y_{\rm mode}$} &
\colhead{AR}  \\
\colhead{} &
\colhead{} &
\colhead{} &
\colhead{} &
\colhead{} &
\colhead{} &
\colhead{(d)} &
\colhead{($10^{-9}$)} &
\colhead{}
}
\startdata
J0537$-$6910 & 66 & 0.498 & 0.056 & $7.6\times10^{4}$  & 0.288 & 156  & 194  & 2.8 \\
J1740$-$3015 & 39 & 0.545 & 0.034 & $9.2\times10^{5}$  & 0.179 & 156  & 65   & 1.8 \\
J1341$-$6220 & 32 & 0.561 & 0.053 & $7.3\times10^{5}$  & 0.156 & 234  & 194  & 4.9 \\
Crab         & 31 & 0.567 & 0.082 & $1.9\times10^{5}$  & 0.161 & 312  & 0    & 8.9 \\
Vela         & 25 & 0.589 & 0.052 & $3.1\times10^{6}$  & 0.120 & 896  & 2393 & 2.6 \\
J0631$+$1036 & 16 & 0.637 & 0.034 & $1.6\times10^{6}$  & 0.188 & 273  & 0    & 1.4 \\
J1801$-$2304 & 14 & 0.652 & 0.052 & $8.0\times10^{5}$  & 0.214 & 507  & 65   & 2.5 \\
J1413$-$6141 & 13 & 0.661 & 0.041 & $1.5\times10^{6}$  & 0.154 & 351  & 323  & 2.1 \\
J1825$-$0935 & 13 & 0.661 & 0.051 & $1.1\times10^{5}$  & 0.154 & 390  & 0    & 31.0 \\
J1048$-$5832 & 12 & 0.671 & 0.057 & $8.3\times10^{6}$  & 0.182 & 468  & 65   & 1.8 \\
\cutinhead{Pulsars with $6 \le N_{\rm g} < 10$}
J1737$-$3137 &  9 & 0.707 & 0.047 & $7.9\times10^{6}$  & 0.111 & 779  & 259  & 2.2 \\
J0205$+$6449 &  8 & 0.723 & 0.048 & $5.0\times10^{6}$  & 0.125 & 662  & 259  & 2.7 \\
J0742$-$2822 &  8 & 0.723 & 0.032 & $7.6\times10^{6}$  & 0.000 & 896  & 0    & 1.9 \\
J2229$+$6114 &  8 & 0.723 & 0.041 & $1.4\times10^{6}$  & 0.125 & 779  & 194  & 1.3 \\
J1801$-$2451 &  8 & 0.742 & 0.045 & $6.4\times10^{6}$  & 0.143 & 1637 & 2587 & 1.9 \\
J1420$-$6048 &  7 & 0.742 & 0.042 & $8.1\times10^{5}$  & 0.000 & 974  & 1293 & 5.2 \\
J1731$-$4744 &  7 & 0.742 & 0.034 & $3.7\times10^{6}$  & 0.000 & 1052 & 0    & 2.1 \\
J1814$-$1744 &  7 & 0.742 & 0.037 & $5.3\times10^{4}$  & 0.000 & 1169 & 0    & 43.5 \\
J1826$-$1334 &  7 & 0.742 & 0.051 & $1.2\times10^{7}$  & 0.286 & 2260 & 2716 & 1.2 \\
J1841$-$0524 &  7 & 0.742 & 0.054 & $1.5\times10^{6}$  & 0.143 & 623  & 970  & 1.3 \\
J1023$-$5746 &  6 & 0.765 & 0.044 & $2.0\times10^{6}$  & 0.000 & 546  & 3040 & 7.9 \\
J1645$-$0317 &  6 & 0.765 & 0.000 & $9.6\times10^{4}$  & 0.000 & 1676 & 0    & 30.4 \\
\enddata
\tablecomments{
Columns: $N_{\rm g}$, sample size after cleaning; $h$, geometric-mean bandwidth in whitened space; $\lambda_q$, HDR density threshold; $A_{\rm HDR}$, HDR area in physical coordinates; $f_{\rm out}$, fraction of events outside the HDR; $(x_{\rm mode}, y_{\rm mode})$, density-mode location; AR, anisotropy ratio defined as the square root of the eigenvalue ratio of the sample covariance matrix.
}
\end{deluxetable*}

The KDE--HDR morphological parameters for all pulsars in our sample are summarized in Table~\ref{tab:hdr_params}. For each source, we list the number of glitches $N_{\rm g}$, the KDE bandwidth in whitened space, and the derived HDR descriptors: the HDR area $A_{\rm HDR}$, the anisotropy ratio AR, the out-of-HDR fraction $f_{\rm out}$, and the density-mode location in the $\Delta t_{-}$--$\Delta\nu/\nu$ plane. The HDR areas span more than two orders of magnitude, from $5.3\times10^{4}$ to $1.2\times10^{7}$. Anisotropy ratios range from values close to unity all the way up to $\mathrm{AR}\approx44$. This wide spread confirms the diversity of morphologies present in the sample---from compact and nearly isotropic HDRs to highly elongated structures with substantial low-density extensions.

What these per-source results show is that the glitch occupancy of individual pulsars can generally be described by smooth, predominantly unimodal density fields. The global geometry of these fields, however, varies considerably from one source to another. A natural question then arises: do these geometric differences organize into a small number of population-level classes? We address this question in the following subsection, where the per-source KDE--HDR descriptors are used as input to a clustering analysis.

\subsection{Population-Level Morphological Classes}
\label{sec:population_classes}

\begin{deluxetable*}{lcccccc}[htbp]
\tablecaption{Robustness of classification under single-feature ablation.\label{tab:class_robustness}}
\tablewidth{0pt}
\tablehead{
\colhead{Source} &
\colhead{HC} &
\colhead{no $f_{\rm out}$} &
\colhead{no $\log{\rm AR}$} &
\colhead{no $\log A_{\rm HDR}$} &
\colhead{Stability} &
\colhead{Status}
}
\startdata
J0537$-$6910 & 1 & 2 & 1 & 1 & 0.344 & Complex$^{\dagger}$ \\
J1341$-$6220 & 1 & 2 & 1 & 1 & 0.401 & Complex$^{\dagger}$ \\
J1420$-$6048 & 2 & 2 & 2 & 2 & 0.849 & Robust \\
J1023$-$5746 & 2 & 2 & 1 & 2 & 0.850 & Robust \\
J1801$-$2451 & 2 & 1 & 1 & 2 & 0.855 & Robust \\
J1801$-$2304 & 2 & 1 & 1 & 1 & 0.870 & Robust \\
J1731$-$4744 & 2 & 1 & 2 & 2 & 0.880 & Robust \\
J0742$-$2822 & 2 & 1 & 2 & 2 & 0.881 & Robust \\
Crab         & 1 & 2 & 1 & 1 & 0.886 & Robust \\
J1413$-$6141 & 2 & 1 & 1 & 1 & 0.895 & Robust \\
J1841$-$0524 & 2 & 1 & 1 & 1 & 0.895 & Robust \\
Vela         & 2 & 1 & 2 & 1 & 0.896 & Robust \\
J0205$+$6449 & 2 & 1 & 2 & 1 & 0.898 & Robust \\
J1826$-$1334 & 2 & 1 & 1 & 1 & 0.898 & Robust \\
J0631$+$1036 & 2 & 1 & 1 & 1 & 0.899 & Robust \\
J2229$+$6114 & 2 & 1 & 1 & 1 & 0.900 & Robust \\
J1814$-$1744 & 1 & 2 & 2 & 2 & 0.901 & Robust \\
J1048$-$5832 & 2 & 1 & 2 & 1 & 0.901 & Robust \\
J1825$-$0935 & 1 & 2 & 1 & 2 & 0.917 & Robust \\
J1737$-$3137 & 2 & 1 & 2 & 1 & 0.917 & Robust \\
J1740$-$3015 & 2 & 1 & 1 & 1 & 0.921 & Robust \\
J1645$-$0317 & 1 & 2 & 1 & 2 & 0.923 & Robust\\
\enddata
\tablecomments{
HC denotes the baseline cluster label using all three features. Subsequent columns show the label assigned when the specified feature is excluded. The Stability column gives the fraction of bootstrap realizations in which the source retains its baseline cluster assignment. Sources with stability $<0.8$ are flagged as Complex.
}
\end{deluxetable*}

\begin{deluxetable*}{ccccc}[htbp]
\tablecaption{Sensitivity of the two-class partition to KDE bandwidth ($h_{\rm mult}$) and HDR threshold ($q$).\label{tab:param_robustness}}
\tablewidth{0pt}
\tablehead{
\colhead{$h_{\rm mult}$} &
\colhead{$q$} &
\colhead{ARI} &
\colhead{NMI} &
\colhead{$f_{\rm flip}$}
}
\startdata
0.5 & 0.50 & 0.176 & 0.135 & 0.273 \\
1.0 & 0.50 & 1.000 & 1.000 & 0.000 \\
2.0 & 0.50 & 1.000 & 1.000 & 0.000 \\
0.5 & 0.68 & 0.152 & 0.208 & 0.727 \\
1.0 & 0.68 & 1.000 & 1.000 & 0.000 \\
2.0 & 0.68 & 1.000 & 1.000 & 0.000 \\
0.5 & 0.90 & 0.541 & 0.459 & 0.909 \\
1.0 & 0.90 & 0.404 & 0.361 & 0.136 \\
2.0 & 0.90 & 0.540 & 0.459 & 0.909 \\
\enddata
\tablecomments{
Agreement with the baseline classification ($h_{\rm mult}=1.0$, $q=0.68$) is quantified by the adjusted Rand index (ARI) and normalized mutual information (NMI). $f_{\rm flip}$ denotes the fraction of sources that change class labels relative to the baseline. All tests use $N_{\rm src}=22$ sources.
}
\end{deluxetable*}

\begin{figure*}[ht!]
\plotone{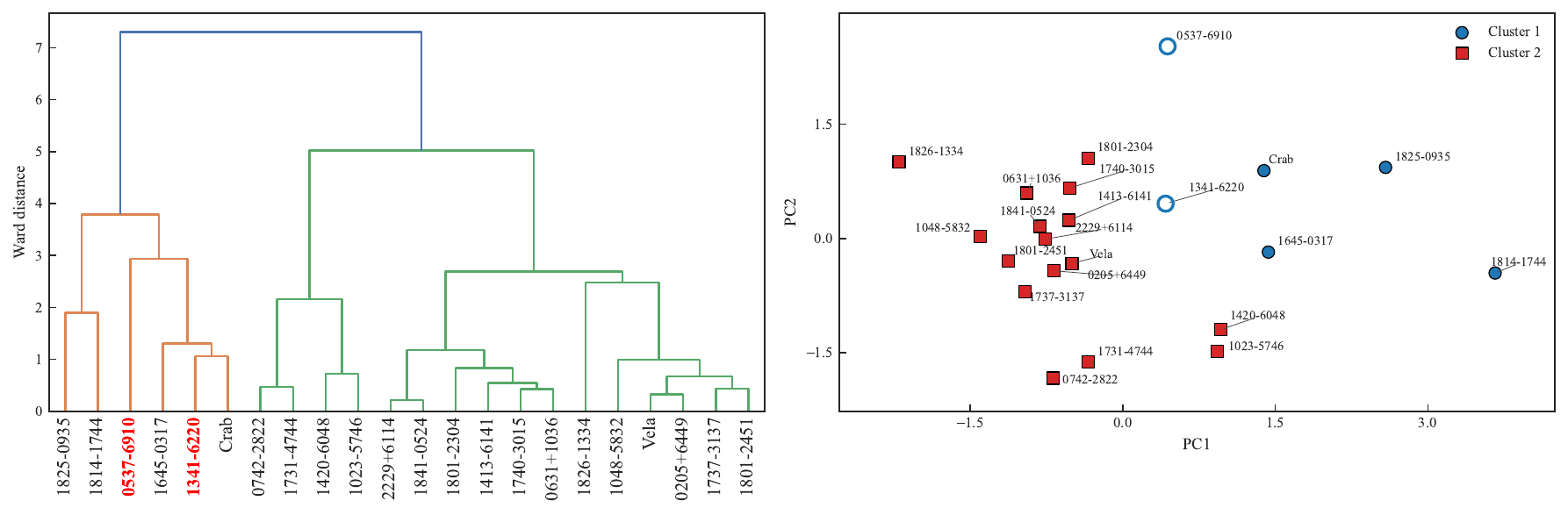}
\caption{Hierarchical clustering and principal component analysis (PCA) of the pulsar sample based on KDE--HDR features ($A_{\rm HDR}$, AR, $f_{\rm out}$). \textbf{Left:} Dendrogram using Ward linkage, showing a primary bifurcation into two classes. \textbf{Right:} Projection of sources onto the first two principal components (PC1, PC2). Colors and symbols correspond to the clusters identified in the dendrogram. The two ``Complex'' sources (PSR~J0537$-$6910 and PSR~J1341$-$6220, marked with open circles) occupy intermediate positions in the PCA plane, consistent with their lower stability scores.}
\label{fig:dendrogram}
\end{figure*}

Hierarchical clustering with Ward linkage (Section~\ref{sec:methods_cluster}) was applied to the KDE--HDR feature vectors $(\log A_{\rm HDR}, \log {\rm AR}, f_{\rm out})$. The resulting dendrogram is shown in Fig.~\ref{fig:dendrogram}. A clear primary bifurcation separates the pulsars into two groups. This dominant split is driven by a joint variation in core spread and tail contribution: one group occupies a region of the feature space defined by compact, coherent cores, while the other is shifted toward larger effective core extent and more prominent peripheral occupancy.

Cluster~1 contains six sources (Figure~\ref{fig:kde_c1}). Four are robust Compact members, while PSR~J0537$-$6910 and PSR~J1341$-$6220 retain the baseline Cluster~1 assignment but are flagged as Complex because of their lower bootstrap stability. Cluster~2 contains the remaining 16 sources (Figure~\ref{fig:kde_c2}) and is referred to as the Extended class. Relative to the robust Compact members, this branch is shifted toward larger HDR areas, lower anisotropy, and/or stronger out-of-HDR occupancy. These labels refer to positions in the three-dimensional descriptor space rather than to visual judgments based on individual KDE panels.

A principal component analysis was performed on the KDE--HDR feature vectors. The first two components account for most of the variance in the three-dimensional feature space. PC1 alone explains 72.4\%, and PC2 adds another 19.1\%, bringing the cumulative variance to 91.5\%. Because the sign of a PCA axis is arbitrary, we interpret the components using the sign convention plotted in Figure~\ref{fig:dendrogram}. In this convention, larger positive PC1 values correspond to smaller $A_{\rm HDR}$ and larger AR, whereas negative PC1 values correspond to broader HDR occupancy. PC1 therefore separates compact, anisotropic high-density regions from broader and more diffuse occupancy fields. PC2 captures the remaining morphological contrast: it mainly traces increasing $f_{\rm out}$, with a secondary negative contribution from $\log A_{\rm HDR}$ and negligible contribution from $\log {\rm AR}$. We therefore treat PC1 and PC2 as summary axes of the standardized KDE--HDR descriptor space rather than as one-to-one proxies for individual observables.

%%A principal component analysis was performed on the KDE--HDR feature vectors. The first two components account for most of the variance in the three-dimensional feature space. PC1 alone explains 72.4\%, and PC2 adds another 19.1\%, bringing the cumulative variance to 91.5\%. Physically, PC1 reflects the overall extent of glitch occupancy. It loads positively on both $\log A_{\rm HDR}$ and $f_{\rm out}$, so it increases with the size of the high-density core and with the degree of peripheral occupation. PC2, on the other hand, captures morphological variation at a given extent. It is dominated by contrasts in anisotropy and orientation, and it separates sources that have similar occupancy scales but different internal shapes.

%%We noted that this finer partition into three groups is sensitive to feature removal---especially whether $f_{\rm out}$ is included or not. This suggests that the evidence for three fully distinct populations is considerably weaker than for the primary two-class structure.

To examine whether the inferred morphology is primarily driven by the characteristic glitch amplitude, we defined the typical amplitude of each source as the median of $\log_{10}(\Delta\nu/\nu)$. We calculated its Spearman rank correlations with $\log A_{\rm HDR}$ and PC1. Uncertainties were estimated using nonparametric bootstrap confidence intervals. 

The typical glitch amplitude shows a moderate positive correlation with $\log A_{\rm HDR}$, with $\rho_{\rm s}=0.414$ and $p=0.0555$. However, the bootstrap 95\% confidence interval, $[-0.034,\,0.772]$, includes zero. Its correlation with PC1 is weaker and also statistically inconclusive: $\rho_{\rm s}=0.335$, $p=0.128$, with a bootstrap 95\% confidence interval of $[-0.152,\,0.713]$ under the PC1 sign convention adopted in Figure~\ref{fig:dendrogram}. 

The robustness of the two-class classification was assessed in three complementary ways. First, label stability was quantified under bootstrap resampling by computing a per-source stability score. Most sources retain their baseline cluster assignment in more than 85\% of realizations. This indicates that the classification is not driven by sampling noise.

Second, a one-feature ablation test was conducted. The hierarchical clustering was repeated after removing each KDE--HDR feature in turn. Table~\ref{tab:class_robustness} shows that most sources preserve their original labels regardless of which feature is dropped. When label changes do occur, they tend to involve sources with low stability scores, which suggests that sensitivity to feature choice reflects intrinsic classification uncertainty rather than dependence on a single parameter.

Third, the two key methodological choices---the KDE bandwidth and the HDR probability threshold---were systematically varied. We rescaled the Silverman bandwidth by a factor $h_{\rm mult} \in \{0.5,\,1.0,\,2.0\}$ and tested HDR probability masses $q \in \{0.50,\,0.68,\,0.90\}$. This grid spans a broad but physically reasonable range. For each combination, the KDE--HDR features were recomputed and the clustering was repeated for all 22 sources. The resulting classifications were compared to the baseline ($h_{\rm mult}=1.0$, $q=0.68$) using the adjusted Rand index (ARI) and normalized mutual information (NMI), both of which are invariant to label permutation. The results are summarized in Table~\ref{tab:param_robustness}.

At the fiducial HDR mass $q=0.68$, the Compact/Extended assignments are exactly reproduced when $h_{\rm mult}=1.0$ or $2.0$ (ARI = NMI = 1.0). This demonstrates that the two-class structure is robust to moderate changes in the smoothing. Reducing the bandwidth to $h_{\rm mult}=0.5$, however, leads to a sharp drop in agreement (ARI $\simeq 0.15$, NMI $\simeq 0.21$). Overly local density estimates evidently introduce instability into the classification.

When the more restrictive threshold $q=0.50$ is used, perfect agreement with the baseline is again obtained for $h_{\rm mult}=1.0$ and $2.0$ (ARI = NMI = 1.0). This means that the core structure by itself is sufficient to recover the two classes. At $h_{\rm mult}=0.5$, agreement drops substantially once more (ARI $\simeq 0.18$, NMI $\simeq 0.14$), consistent with the sensitivity to undersmoothing noted above.

For the more inclusive threshold $q=0.90$, agreement decreases across all bandwidths (ARI $\simeq 0.40$--0.54; NMI $\simeq 0.36$--0.46). Large flip fractions accompany the results at $h_{\rm mult}=0.5$ and $2.0$. Including the low-density tails evidently reduces the contrast between Compact and Extended morphologies, making the classification more sensitive to boundary fluctuations.

Taken together, these tests indicate that the two-class structure is robust within a well-defined region of parameter space---particularly for $q = 0.68$ and moderate bandwidths. When the parameters are pushed outside this regime, either through excessive localization or through inclusion of diffuse tails, the stability of the inferred classification degrades.

One final point deserves mention. The raw flip fraction $f_{\rm flip}$ (Table~\ref{tab:param_robustness}) is by design a stringent per-source measure. It can appear large even when ARI and NMI point to substantial partition similarity. This is because ARI and NMI capture global agreement of the partition structure, whereas $f_{\rm flip}$ counts every per-source label change equally. In sum, the Compact/Extended split does not appear to be an artifact of finely tuned bandwidth or HDR threshold choices. Rather, it reflects a genuine separation in the KDE--HDR feature space, with sensitivity limited to a small number of borderline sources under extreme parameter settings.

\subsection{Physical Drivers of Glitch Morphology}
\label{sec:physical}

We next investigated the physical drivers behind the Compact/Extended morphological classification by examining how the KDE--HDR features relate to pulsar physical parameters. Spearman rank correlation coefficients were computed separately for the two morphological classes, which allows class-dependent trends to be identified that might otherwise be masked in a pooled analysis. We found that $f_{\rm out}$ and $\log{\rm AR}$ show strong sign reversals in their correlations with characteristic timescales ($\log\langle\Delta t\rangle$ and $\log(\langle\Delta t\rangle|\dot\nu|)$). The differences $\Delta\rho$ between the two classes are large. This means that the same physical variable can play qualitatively different roles in shaping glitch occupancy geometry, depending on which morphological regime the pulsar belongs to.

These class-dependent correlations motivated us to build a logistic regression model. The goal was to quantify how well physical parameters can discriminate between the Compact and Extended morphologies. Six predictors were used: $\log \langle \Delta t \rangle$ (mean waiting time), $\log \nu$ (spin frequency), $\log |\dot{\nu}|$ (spin-down rate), $\log ( \langle \Delta t \rangle |\dot{\nu}|)$ (accumulated spin-down scale), $\log \tau_c$ (characteristic age), and $\log B$ (surface magnetic field). The resulting model achieves an area under the receiver operating characteristic curve of $\mathrm{AUC}=0.88$, indicating a statistically significant association between spin-down properties and glitch distribution geometry.

To evaluate the robustness of this result, we performed leave-one-out cross-validation (LOO-CV). In each fold, the classifier was trained on 21 sources and tested on the remaining one. The cross-validated performance is $\mathrm{AUC_{CV}} = 0.59$---substantially lower than the in-sample value. This gap indicates that the logistic model is susceptible to overfitting, which is not surprising given the small sample ($N_{\mathrm{src}} = 22$ with six predictors). With only six members in the Compact class, removing a single source can shift the decision boundary appreciably. The regression coefficients listed in Table~\ref{tab:phys_combined} should therefore be read as indicating the \emph{relative importance hierarchy} among predictors, with $\log\langle\Delta t\rangle$ clearly dominant, rather than as precise discriminative boundaries. Despite the limited generalization, it is worth noting that the qualitative ordering of predictor importance is stable: $\log\langle\Delta t\rangle$ retains the largest coefficient magnitude in all 22 LOO-CV folds.

The fitted coefficients (Table~\ref{tab:phys_combined}) confirm that timescale-related parameters dominate the discrimination. The mean waiting time $\log \langle \Delta t \rangle$ carries the largest positive coefficient, followed by $\log ( \langle \Delta t \rangle|\dot{\nu}|)$ and $\log |\dot{\nu}|$. The parameters $\log B$ and $\log \tau_c$ enter with smaller negative weights. This hierarchy suggests that it is the temporal organization of glitch activity, and the associated stress accumulation timescale, that primarily regulates the Compact/Extended dichotomy---rather than magnetic field strength or instantaneous spin frequency.

\begin{deluxetable*}{llcccccc}[htbp]
\tablecaption{Correlations between pulsar physical properties and KDE--HDR morphological features.\label{tab:phys_combined}}
\tablewidth{0pt}
\tablehead{
\colhead{Physical parameter} &
\colhead{KDE--HDR feature} &
\colhead{$\rho_{\rm C}$} &
\colhead{$\rho_{\rm E}$} &
\colhead{$\Delta\rho$} &
\colhead{Logistic coef}
}
\startdata
$\log B$ & $f_{\rm out}$   & $-0.900$ &  0.151 & $-1.051$ & $-0.237$ \\
         & $\log A$        & $-0.100$ & $-0.262$ &  0.162 & \nodata \\
         & $\log {\rm AR}$ &  0.700 &  0.056 &  0.644 & \nodata \\
\hline
$\log |\dot{\nu}|$ & $f_{\rm out}$   &  0.900 & $-0.093$ &  0.993 &  0.145 \\
                   & $\log A$        &  0.500 & $-0.035$ &  0.535 & \nodata \\
                   & $\log {\rm AR}$ & $-0.700$ &  0.229 & $-0.929$ & \nodata \\
\hline
$\log \tau_c$ & $f_{\rm out}$   & $-0.800$ &  0.148 & $-0.948$ & $-0.070$ \\
              & $\log A$        & $-0.300$ &  0.153 & $-0.453$ & \nodata \\
              & $\log {\rm AR}$ &  0.600 & $-0.300$ &  0.900 & \nodata \\
\hline
$\log \langle \Delta t \rangle$ & $f_{\rm out}$   & $-0.500$ & $-0.156$ & $-0.344$ &  1.531 \\
                                & $\log A$        &  0.100 &  0.576 & $-0.476$ & \nodata \\
                                & $\log {\rm AR}$ &  0.700 & $-0.115$ &  0.815 & \nodata \\
\hline
$\log (\langle \Delta t \rangle|\dot{\nu}|)$ & $f_{\rm out}$   & $-1.000$ &  0.079 & $-1.079$ &  0.224 \\
                                              & $\log A$        & $-0.300$ &  0.129 & $-0.429$ & \nodata \\
                                              & $\log {\rm AR}$ &  0.900 & $-0.279$ &  1.179 & \nodata \\
\hline
$\log \nu$ & $f_{\rm out}$   &  1.000 & $-0.119$ &  1.119 &  0.202 \\
           & $\log A$        &  0.300 &  0.079 &  0.221 & \nodata \\
           & $\log {\rm AR}$ & $-0.900$ &  0.129 & $-1.029$ & \nodata \\
\enddata
\tablecomments{
Columns 3--5 show Spearman rank correlation coefficients ($\rho$) calculated separately for the Compact (C) and Extended (E) classes, along with the difference $\Delta\rho = \rho_{\rm C} - \rho_{\rm E}$. The final column lists the coefficients from the logistic regression model; positive values indicate a higher probability of belonging to the Extended class.
}
\end{deluxetable*}

\section{Discussion} 
\label{sec:discussion}

\subsection{Physical Interpretation of the Two Morphological Classes}
\label{sec:disc_physics}

The two-class separation seen in Figure~\ref{fig:dendrogram} can be interpreted naturally within the superfluid angular-momentum transfer framework \citep{anderson_pulsar_1975,alpar_vortex_1984,haskell_models_2015,Antonopoulou2022}. In the standard vortex-unpinning picture, angular momentum builds up in the crustal superfluid at a rate set by the spin-down torque. When the velocity lag between the superfluid and the crust grows large enough to cross a critical threshold, vortices unpin collectively and transfer angular momentum to the crust, producing the observed glitch. The morphological signature that a glitch leaves in the $(\Delta t_{-},\,\Delta\nu/\nu)$ plane depends on how thoroughly the reservoir is drained in each event.

Consider first the case in which glitches are triggered by a deterministic critical-lag threshold and the reservoir is nearly emptied each time. Under these conditions, the waiting time and the resulting amplitude should be tightly coupled. In the $(\Delta t_{-},\,\Delta\nu/\nu)$ plane, this would appear as a linear ridge or a concentrated core (e.g., Figure~\ref{fig:kde_c1}). Our Compact class displays exactly this kind of behavior---tightly confined along a preferred direction. On the other hand, stochastic avalanche models predict that only a fraction of the reservoir is tapped during any single event. The resulting distribution of sizes and intervals is broader and less correlated. This is what we observe in the Extended class, which shows a larger $A_{\rm HDR}$ and a more isotropic occupation of parameter space. If the Extended class does indeed reflect partial depletion, then the observed glitch coupling constant $G$ will underestimate the total superfluid moment of inertia that is actually available.

Two sources deserve special attention. PSR~J0537$-$6910 and PSR~J1341$-$6220 were both flagged as ``Complex'' in Table~\ref{tab:class_robustness}. They are prolific glitchers ($N_{\rm g}=66$ and 32, respectively) and they sit at intermediate positions in the PCA plane (Figure~\ref{fig:dendrogram}, right panel). Their bootstrap stability scores fall well below the 0.8 threshold, at 0.34 and 0.40.

The KDE morphologies of these two sources show features of both classes. PSR~J0537$-$6910, for instance, has the second smallest HDR area in the sample ($7.6\times10^{4}$) and a moderate anisotropy ($\mathrm{AR}=2.8$). On these grounds it would belong to the Compact class. Yet its out-of-HDR fraction ($f_{\rm out}=0.29$) is the highest among all 22 pulsars---a trait that is much more characteristic of the Extended class. This tension has a natural physical reading. The well-known strong correlation between glitch size and forward waiting time in this source \citep{middleditch_predicting_2006, Antonopoulou2018MNRAS, Gugercinoglu2026} points to near-complete reservoir depletion in each cycle, which is a Compact-class signature. The elevated $f_{\rm out}$, however, may reflect occasional partial-depletion events that are superposed on the dominant deterministic sequence.

PSR~J1341$-$6220 presents a similar hybrid character. It has a single dominant density core, but this core is stretched along the diagonal direction, and the peripheral occupancy is moderate ($f_{\rm out}=0.16$). \citet{zhu_glitches_2025} found phase-dependent variations in $G$ for this pulsar, consistent with a reservoir that alternates between near-complete and partial depletion.

As glitch catalogs continue to grow, it will be important to track whether these Complex sources drift toward one class or the other, or whether they instead stabilize as a genuine intermediate category. This would provide a direct observational test of whether the Compact/Extended dichotomy is truly discrete or whether it reflects the endpoints of a continuum.

That $\langle\Delta t_{-}\rangle$ emerges as the strongest logistic-regression predictor of class membership (Section~\ref{sec:physical}) is consistent with the result of \citet{Millhouse2022MNRAS}, who showed that the glitch rate scales primarily with characteristic age rather than with spin frequency or magnetic-field strength. Together, these findings reinforce the view that timescale-related quantities---not instantaneous rotational parameters---are the main regulators of glitch-activity morphology.

\subsection{Comparison with Previous Classification Schemes}
\label{sec:disc_compare}

For many years, the community has relied on the qualitative ``Vela-like'' versus ``Crab-like'' dichotomy to distinguish large, quasi-periodic glitchers from small, irregular ones \citep{flanagan_rapid_1990, espinoza_study_2011}. Our Compact/Extended classes overlap partially with this traditional division, but the two are not equivalent. Vela itself falls into the Extended class, whereas the Crab is assigned to the Compact class (Figure~\ref{fig:dendrogram}). This assignment reflects the joint waiting-time--amplitude morphology, not amplitude alone.

A four-class taxonomy was recently proposed by \citet{zhu_glitches_2025}, based on the visual appearance of glitch-size bar charts and the cumulative angular-momentum evolution. Their scheme identifies a linear relationship between the glitch cluster period $P_c$ and characteristic age, which provides a useful physical timescale for glitch recurrence. In our framework, the mean backward waiting time $\langle\Delta t_{-}\rangle$ serves as the physical analog of their cluster period. Their classes 2--4 are distributed across both of our morphological classes, which highlights a key difference in approach: our method extracts the Compact/Extended boundary from a reproducible numerical criterion applied to the full joint two-dimensional distribution, rather than from one-dimensional sequences. Establishing a direct numerical correspondence between our geometric features and their cluster period $P_c$ would be a valuable direction for future work, and could help to unify these two complementary approaches.

\citet{Fuentes2019} carried out a systematic analysis of glitch-size and waiting-time distributions for eight prolific pulsars. They identified characteristic scales in the large-glitch populations of Vela and PSR~J0537$-$6910, while finding Poisson-like behavior in other sources. Our KDE--HDR framework goes further by operating in the full two-dimensional $(\Delta t_{-},\,\Delta\nu/\nu)$ plane. This allows joint morphological features---core anisotropy, peripheral occupancy, and so on---to be captured in a way that is not possible from separate one-dimensional analyses.

An interesting related result was reported by \citet{eya_distributions_2019}, who showed that the inter-glitch time intervals for small-size ($\Delta\nu/\nu < 10^{-7}$) and large-size ($\Delta\nu/\nu \ge 10^{-7}$) glitches are drawn from statistically indistinguishable distributions. This observation fits well with our finding that the Compact/Extended separation is not driven by whether a pulsar produces large or small glitches per se. What matters instead is how the \emph{joint} distribution of sizes and intervals concentrates or spreads in the occupancy plane. That the waiting-time structure contributes independent discriminative information is confirmed by the logistic regression (Section~\ref{sec:physical}), where $\log\langle\Delta t_{-}\rangle$ stands out as the strongest single predictor of class membership.

\subsection{The Role of the Backward Waiting Time}
\label{sec:disc_dt}

The logistic regression analysis (Table~\ref{tab:phys_combined}) identifies $\langle\Delta t_{-}\rangle$ as the strongest single predictor of morphological class. Its coefficient is 1.531, which is substantially larger than those of $\log\nu$ (0.202) and $\log B$ ($-0.237$). 
This result is physically reasonable, as the pre-glitch accumulation timescale directly governs how much angular momentum can be stored in the superfluid reservoir before the next event \citep{melatos_avalanche_2008, howitt_nonparametric_2018}.

The class-dependent Spearman correlations (Table~\ref{tab:phys_combined}) add further detail. In the Compact class, $\log\langle\Delta t_{-}\rangle$ and AR are correlated ($\rho_C = 0.700$). This means that pulsars with longer average accumulation times tend to develop more elongated and structured HDR cores. If a pulsar systematically depletes its reservoir at each glitch, the interpretation is clear: a longer accumulation time allows a larger critical lag to develop. This enforces a tighter deterministic coupling between the pre-glitch waiting time and the resulting amplitude, stretching the occupancy core along a preferred axis. In the Extended class, by contrast, we found that this correlation essentially vanishes ($\rho_E = -0.115$). The sign reversal ($\Delta\rho = 0.815$) implies that when glitches are driven by partial, avalanche-like releases, the accumulation timescale sets the overall glitch rate but does not impose a preferred directional coupling between size and waiting time.

It is worth noting that our analysis relies on the backward waiting time $\Delta t_{-}$, which conditions each event on the pre-glitch state. The forward waiting time $\Delta t_{+}$, however, also carries important physical information. To evaluate the effect of this choice, we repeated the full KDE--HDR analysis using $\Delta t_{+}$ (see Appendix~\ref{app:forward_classification}).

The same pipeline applied to $\Delta t_{+}$ does produce a two-class separation. The classification stability, however, is noticeably reduced. The stability metrics show a broader distribution than in the backward-interval case. Moreover, the logistic regression coefficients exhibit a weaker hierarchical structure: $\log \Delta t$ remains an important predictor, but it no longer dominates the regression as clearly as it does in the backward analysis.

This difference can be understood in terms of the distinct physical roles played by the two intervals. The backward interval traces the pre-glitch accumulation of stress and is therefore directly tied to the angular-momentum reservoir. The forward interval, on the other hand, is shaped by post-glitch recovery processes as well as stochastic triggering, both of which introduce additional variability. As a consequence, the correlations observed with $\Delta t_{+}$ are less coherent.

The case of PSR~J0537$-$6910 illustrates the point well \citep{middleditch_predicting_2006, Antonopoulou2018MNRAS, Ferdman2018ApJ}. The strong correlation between glitch size and waiting time that characterizes this source is known to depend on whether the backward or forward interval is used. In our analysis, the forward-interval classification yields significantly lower stability for this pulsar. This observation supports the interpretation that the morphological structure uncovered in this work is primarily shaped by pre-glitch accumulation rather than by post-glitch evolution.

\subsection{Implications for the Angular-Momentum Reservoir}
\label{sec:disc_reservoir}

The glitch coupling constant $G$ is defined as the ratio of cumulative glitch spin-up to the secular spin-down measured over the same time span. It provides a macroscopic estimate of the fractional stellar moment of inertia that participates in glitches \citep{link_MOI_1999, lyne_statistical_2000}. Through long-term pulsar timing, empirical values of $G$ have been established for a number of prolific sources. For Vela, roughly 1.6\% of the spin-down is reversed on average by glitches ($G \approx 0.016$; \citealt{espinoza_study_2011, Fuentes2019, ho_pinning_2015}). This is a fairly typical value; many rotation-powered pulsars have $G$ in the range $0.01$--$0.02$ \citep{lyne_statistical_2000, espinoza_study_2011}. Recent observations have also shown that Vela glitches can be accompanied by magnetospheric perturbations \citep{palfreyman_alteration_2018, ashton_flickering_2020}, and theoretical modeling suggests that crustal failure---a starquake---can act as the trigger for vortex unpinning \citep{Bransgrove2020}. If this is the case, the angular-momentum release can occur before the critical lag is reached globally, leading to only partial depletion of the reservoir.

The Crab pulsar, by contrast, has glitch activity that is orders of magnitude lower, with $G \sim 10^{-5}$ \citep{espinoza_study_2011, Wang2021MNRAS}. Our morphological classification provides a structural framework within which these phenomenological differences can be understood.

For the Compact class, of which the Crab is the clearest example, the glitch distribution is tightly concentrated in the $(\Delta t_{-},\,\Delta\nu/\nu)$ plane. This concentration points to a highly coherent coupling between the pre-glitch accumulation process and the resulting glitch amplitude. The system appears to operate in a relatively deterministic regime, with each event tapping a well-defined fraction of the available reservoir.

It is important to emphasize, however, that the Crab's low glitch activity does not necessarily mean that its total superfluid reservoir is small. Analysis of the 2017 largest Crab glitch within the vortex creep model indicates that the crustal superfluid fraction participating in the event reaches $I_{\rm creep}/I \approx 6\times10^{-3}$ \citep{Gugercinoglu2019}. Furthermore, \citet{Zhu2026} demonstrated that the long-term quasi-periodicity of Crab glitches and the correlations between cluster size and waiting time are consistent with a partial-release mechanism, in which each event taps only a fraction of the stored superfluid angular momentum. This picture aligns naturally with our Compact-class interpretation: the coupling to the observable glitch channel is coherent, but the reservoir itself may be substantially larger than what the long-term $G$ alone would suggest.

In the Extended class, with Vela as the prototype, the dispersed occupancy indicates that the reservoir is only partially tapped at each event. The long-term observed $G$ therefore underestimates the true superfluid participation. This conclusion receives independent support from the phase-dependent $G$ variations found in PSR~J1341$-$6220 by \citet{zhu_glitches_2025}.

Even before partial depletion is taken into account, microphysical entrainment corrections already push the required superfluid fraction upward. \citet{Andersson2012PhRvL} and \citet{Chamel2013PhRvL} showed that nondissipative entrainment coupling increases the effective mass of the superfluid neutrons in the crust. To explain Vela's apparent $G\approx 0.016$, the actual superfluid moment-of-inertia fraction $I_{\rm res}/I_{\rm c}$ must reach at least $\sim 7\%$ once entrainment is included. This requires the participation of superfluid that extends into the stellar outer core \citep{ho_pinning_2015, montoli_role_2020}.

If the avalanche-like, partial-depletion mechanism inferred from the Extended morphology operates on top of the entrainment correction, then the true $I_{\rm res}/I_{\rm c}$ must be strictly greater than even the entrainment-corrected bound. Event-wise comparisons of the fractional moment of inertia required by individual glitches confirm that a non-negligible fraction of observed events cannot be explained by the crustal superfluid alone \citep{Basu2018}. The inferred ratio should therefore be treated as a strict lower limit. Population-level comparisons of $G$, and the consequent constraints on the equation of state of dense matter, should account for morphological class membership. The quantitative framework introduced here makes such a correction feasible.

\subsection{Limitations and Future Directions}
\label{sec:disc_limits}

The present analysis has several limitations that naturally point to avenues for future work. The most important is sample size. With only 22 pulsars, the statistical power of both the classification and the logistic regression is limited. For sources near the minimum event threshold ($N_{\rm g}=6$), the KDE resolution inherently suppresses subtle out-of-core features---this is why many of these sources have $f_{\rm out} \approx 0$ (Table~\ref{tab:hdr_params}). The low ratio of sample size to number of predictors in the regression also means that the exact coefficient magnitudes should be viewed as indicative trends, not definitive bounds. As larger glitch catalogs become available through long-term monitoring campaigns and next-generation facilities such as FAST and the SKA, it will be essential to expand the sample. Expanding the sample will sharpen the morphological class boundaries and allow more robust discriminative models to be trained.

PSR~J0742$-$2822 warrants additional caution when its individual KDE morphology is interpreted. This pulsar exhibits rapid changes in both pulse-profile shape and spin-down rate, which have been associated with magnetospheric state switching and can contribute substantial timing noise \citep{lyne_switched_2010, keith_connection_2013, brook_emission_2016, lower_ubiquity_2025}. Consequently, some of the lowest-amplitude catalogued events may be more difficult to distinguish from state-dependent rotational irregularities than the larger glitches. At the same time, high-cadence timing has confirmed at least one small glitch in this source despite its strong timing noise \citep{basu_observed_2020}; it would therefore be inappropriate to reject all of its small events solely on the basis of amplitude. Our leave-one-source-out stability analysis indicates that the global two-class partition is not controlled by any single pulsar, thus this source-specific ambiguity does not determine the population-level
result. Nevertheless, the KDE--HDR parameters and class assignment of PSR~J0742$-$2822 should be interpreted with greater caution than those of sources whose glitch records are less affected by state switching. Future high-cadence timing and a homogeneous re-evaluation of its low-amplitude events will be needed to refine its individual morphology.

A second limitation is that the KDE--HDR parameters, while providing a reproducible phenomenological summary of glitch occupancy, do not by themselves establish a causal connection to specific internal mechanisms. One promising way forward is through forward modeling. If synthetic glitch catalogs generated from physically motivated models \citep{warszawski_gross-pitaevskii_2011, melatos_size-waiting-time_2018} can reproduce the observed morphological classes, that would considerably strengthen the physical interpretation. In addition, extending the framework to include the forward waiting time ($\Delta t_{+}$) and tracking how morphology evolves over decades would help to clarify whether the two-class structure captures post-glitch relaxation dynamics as well, and whether morphologically ``complex'' pulsars transition between different activity regimes as they age and spin down.

\section{Conclusion}
\label{sec:conclusion}

In this work, we have developed a quantitative framework for classifying pulsar glitch behavior on the basis of morphology in the joint space of 
backward waiting time and relative glitch amplitude. Qualitative taxonomies are replaced by probabilistic descriptors derived from two-dimensional kernel density estimation and highest-density regions (KDE--HDR), which together provide a reproducible summary of glitch occupancy. We applied hierarchical clustering to the geometric parameters ($A_{\rm HDR}$, ${\rm AR}$, $f_{\rm out}$) extracted from 349 glitches across 22 pulsars. The analysis reveals a robust two-class structure. One class is characterized by compact, coherent cores with limited anisotropy. The other exhibits broader distributions with substantial out-of-core contributions. Robustness tests---including bootstrap resampling, feature ablation, and hyperparameter variation---confirm that this separation reflects genuine differences in glitch-occupancy morphology rather than artifacts of sampling or methodological choices.

To explore the physical drivers of these morphological classes, we constructed a logistic regression model using spin and waiting-time parameters as predictors. The model achieves an in-sample classification performance of $\mathrm{AUC}=0.88$. The leave-one-out cross-validated 
value, $\mathrm{AUC}_{\mathrm{CV}} = 0.59$, is considerably lower, reflecting the limited sample size. Nevertheless, the predictor-importance hierarchy remains stable across all folds. Among the predictors examined, the mean backward waiting time $\langle\Delta t_{-}\rangle$ stands out as the most informative, suggesting that the pre-glitch stress-accumulation timescale plays a central role in shaping glitch occupancy. From a physical standpoint, the Compact class appears consistent with near-complete depletion of the angular-momentum reservoir in each glitch cycle, while the Extended class points to partial, avalanche-like releases. As a consequence, short-term measurements of the glitch coupling constant $G$ in Extended-class pulsars may underestimate the true fraction of the crustal superfluid that participates in glitch activity.

\begin{acknowledgments}
We would like to thank the anonymous referee for helpful
suggestions that led to a significant improvement of our work. This work is supported by the Natural Science Foundation of Xinjiang Uygur Autonomous Region (No. 2023D01E20), the Xinjiang Talent Development Fund (No. XJRC-2025-KJ-PY-KJLJ-106), the National Key R\&D Program of China (No. 2022YFA1603104), the Tianshan talents program (No. 2023TSYCTD0013), the National Natural Science Foundation of China (No. 12288102).
Weihua Wang is supported by the Zhejiang Provincial Natural Science Foundation of China under grant No. LQ24A030002. Feifei Kou is supported by the Major Science and Technology Program of Xinjiang Uygur Autonomous Region (2022A03013-2).
\end{acknowledgments}

\begin{contribution}
%%This section gives authors the space to recognize author contributions. The text inside this environment is NOT counted towards the total word quanta. At a minimum, manuscripts are expected to include this text:

W.-T.L. developed the methodology, performed the analysis, and drafted the manuscript.
W.-H.W. contributed to the interpretation.
X.Z. conceived the project, supervised the work and edited the manuscript.
F.-F.K. contributed to the interpretation.

%% But authors are expected to provide more specific details, e.g. 
%%
%%SC was responsible for writing and submitting the manuscript.
%%WWM came up with the initial research concept and edited the manuscript.
%%OTS obtained the funding and edited the manuscript.
%%EBF provided the formal analysis and validation. He also edited the manuscript.
%%GEH Supervised the undergraduates, wrote the software and administers the project github and Zenodo repositories.
%%
%% Authors can use the Contributor Role Taxonomy (CRediT) at
%% for ideas on how write a good statement tailored to their needs.

\end{contribution}


\software{
Python, 
NumPy \citep{harris2020array},
SciPy \citep{2020SciPy-NMeth},
pandas \citep{mckinney-proc-scipy-2010},
Matplotlib \citep{Hunter:2007},
scikit-learn \citep{scikit-learn}
          }


\appendix

\section{Clustering Tendency Diagnostics}
\label{app:hopkins}

The Hopkins statistic ($H$) is used here to test whether the glitch occupancy of an individual pulsar shows statistically significant intrinsic clustering. It is a standard diagnostic for clustering tendency in multivariate point sets, and it is particularly useful for distinguishing genuine clustering from a single continuous distribution that happens to contain smooth gradients.

The procedure is as follows. For a given pulsar with $N$ observed glitches in the logarithmic $(\Delta t_{-},\Delta\nu/\nu)$ plane, denoted $\{\mathbf{x}_i\}_{i=1}^{N}$, we generate a set of $m$ synthetic points $\{\mathbf{u}_j\}_{j=1}^{m}$ uniformly at random within the axis-aligned bounding box of the data. Separately, $m$ observed points $\{\mathbf{x}^{(s)}_j\}_{j=1}^{m}$ are drawn uniformly at random from the original sample. Two sets of nearest-neighbor distances are then computed. The first set, $u_j$, gives the distance from the $j$-th random point to its nearest neighbor among the observed data. The second set, $w_j$, gives the distance from the $j$-th sampled real point to its nearest neighbor in the observed set. The Hopkins statistic is then
\begin{equation}
H = \frac{\sum_{j=1}^{m} u_j}{\sum_{j=1}^{m} u_j + \sum_{j=1}^{m} w_j}.
\label{eq:hopkins}
\end{equation}

Data that are consistent with complete spatial randomness (CSR) produce $H \simeq 0.5$. Values approaching 1 indicate a strong clustering tendency---the synthetic points land in gaps between clusters and therefore have large $u_j$ relative to the $w_j$ of the real points. Values near 0, conversely, suggest a regularly spaced or inhibited point pattern.

We emphasize that the Hopkins statistic is used here strictly as a diagnostic tool. Its role is to test for discrete internal substructure, such as distinct islands of glitch activity. It is \emph{not} used to define the Compact and Extended morphology classes, which are based instead on the global geometric descriptors $A_{\rm HDR}$, AR, and $f_{\rm out}$. Table~\ref{tab:hopkins} lists the Hopkins statistics for all sources that pass our minimum-sample criterion. We interpret these values qualitatively. Sources with $H \approx 0.5$ are consistent with continuous distributions. Those with substantially higher values---Vela being a prominent example---provide evidence for intrinsic multi-modal clustering.

\begin{deluxetable}{lcclcclcc}[htbp]
\tablecaption{Hopkins statistic ($H$) for clustering tendency assessment.
\label{tab:hopkins}}
\tablewidth{0pt}
\tablehead{
\colhead{Source} & \colhead{$N_{\rm g}$} & \colhead{$H$} &
\colhead{Source} & \colhead{$N_{\rm g}$} & \colhead{$H$} &
\colhead{Source} & \colhead{$N_{\rm g}$} & \colhead{$H$}
}
\startdata
J0205$+$6449 &  8 & 0.234 & J0742$-$2822 &  8 & 0.371 & J1413$-$6141 & 13 & 0.532 \\
J0537$-$6910 & 66 & 0.446 & J1023$-$5746 &  6 & 0.282 & J1420$-$6048 &  7 & 0.532 \\
J0631$+$1036 & 16 & 0.436 & J1048$-$5832 & 12 & 0.318 & J1645$-$0317 &  6 & 0.485 \\
J1341$-$6220 & 32 & 0.641 & J1731$-$4744 &  7 & 0.435 & J1737$-$3137 &  9 & 0.441 \\
J1740$-$3015 & 39 & 0.494 & J1801$-$2304 & 14 & 0.612 & J1801$-$2451 &  7 & 0.382 \\
J1814$-$1744 &  7 & 0.898 & J1825$-$0935 & 13 & 0.281 & J1826$-$1334 &  7 & 0.563 \\
J1841$-$0524 &  7 & 0.124 & J2229$+$6114 &  8 & 0.208 & Crab         & 31 & 0.743 \\
Vela         & 25 & 0.658 & \nodata      & \nodata & \nodata & \nodata      & \nodata & \nodata \\
\enddata
\tablecomments{
$H$ is calculated according to Eq.~(\ref{eq:hopkins}). Values near 0.5 imply a continuous distribution, while values approaching 1.0 suggest strong intrinsic clustering.
}
\end{deluxetable}

\section{Forward Waiting-Time Classification}
\label{app:forward_classification}

As a check on the robustness of our morphological classification, we repeated the full KDE--HDR analysis using the forward interval $\Delta t_{+}$ instead of the backward interval. The forward waiting time is defined as the time from a given glitch to the next event. Every other aspect of the analysis pipeline---KDE construction, HDR extraction at fixed probability mass, feature derivation, and hierarchical clustering in the $(\log A, \log {\rm AR}, f_{\rm out})$ space---was kept identical to what is described in Section~\ref{sec:methods_kde}.

When the KDE--HDR framework is applied to the forward interval, a two-class structure emerges that is broadly analogous to the one obtained with the backward interval. The clustering algorithm again splits the sample into two groups, which confirms that the bimodal morphological behavior is not simply an artifact of the choice of temporal direction.

\begin{figure*}[ht!]
\plotone{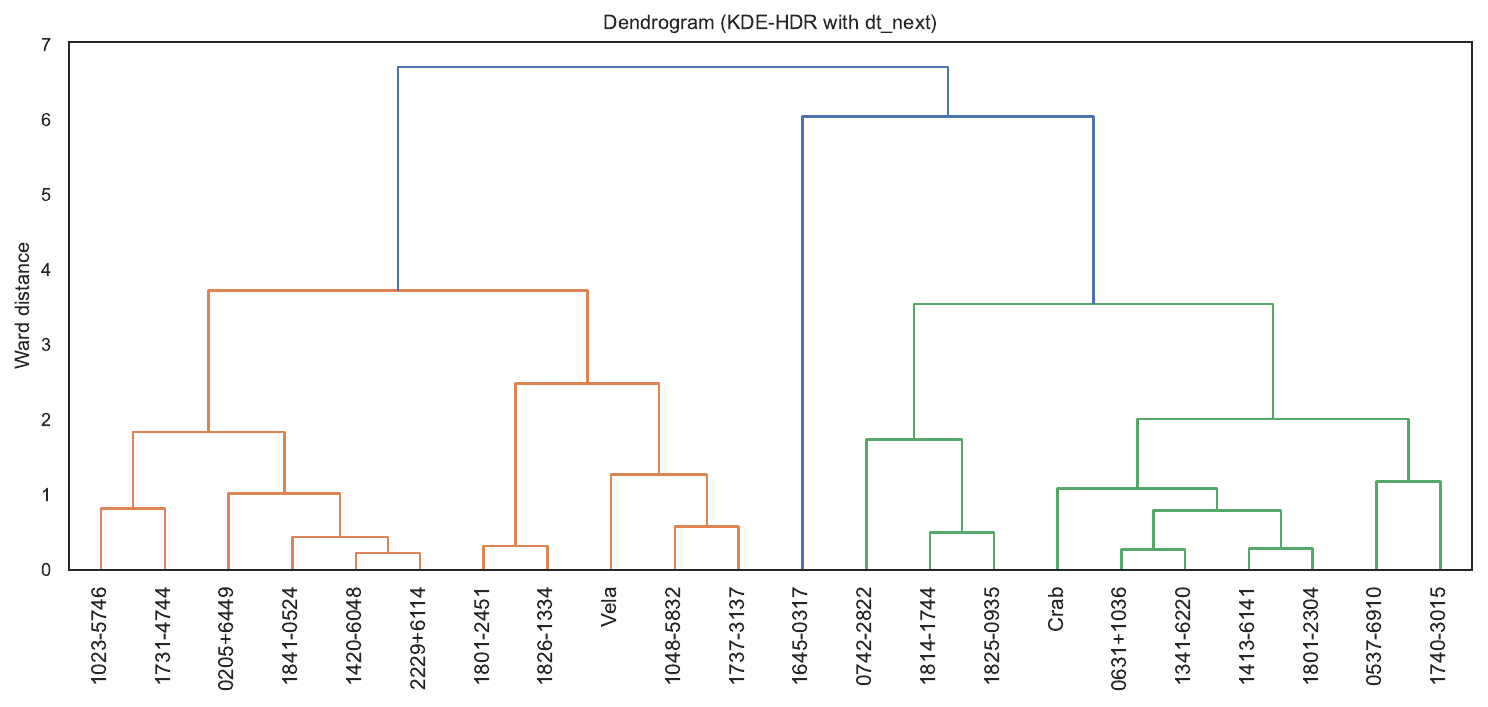}
\caption{Dendrogram of pulsars in the KDE--HDR feature space constructed using the forward waiting time $\Delta t_{+}$. 
The same analysis pipeline as in the backward-interval case was applied. 
Although a two-class structure is recovered, the separation is weaker and less stable, indicating reduced robustness of the forward-interval representation.}
\label{fig:forward_dendrogram}
\end{figure*}

That said, the separation between the two classes is less clean in the forward-interval representation. The KDE--HDR feature distributions show more overlap, and the cluster boundaries are less sharply defined. This can also be seen in the dendrogram (Figure~\ref{fig:forward_dendrogram}), where the linkage distances between clusters are smaller than in the backward-interval case.

We evaluated the stability of the forward-interval classification using the same bootstrap procedure described in Section~\ref{sec:methods_cluster}. The results are summarized in Table~\ref{tab:forward_stability}. Across the board, the forward-interval classification is less stable than its backward-interval counterpart. The HDR extraction failure rate is higher, and key features like $\log A$ and $f_{\rm out}$ show greater dispersion across bootstrap realizations. For several sources, the effective number of successful realizations $B_{\rm eff}$ is noticeably reduced. These findings indicate that the KDE--HDR features derived from $\Delta t_{+}$ are more sensitive to sampling fluctuations, which results in a less reliable characterization of glitch morphology.

We also examined how pulsar physical parameters relate to the forward-interval morphological classes through logistic regression. The outcome differs from the backward-interval case in an instructive way. With the backward interval, $\log \langle \Delta t_{-} \rangle$ clearly dominates the regression. In the forward-interval analysis, by contrast, the coefficients are more evenly distributed. The parameter $\log \langle \Delta t_{+} \rangle$ is still among the leading predictors, but its coefficient is no longer much larger than those of the other physical parameters. This reduced hierarchy suggests that the forward interval does not serve as a primary organizing variable for morphological class. Instead, it appears to contribute in combination with other parameters, pointing to a more complex and less deterministic relationship.

In summary, these results show that the two-class morphological structure does persist when the forward interval is used, but its statistical robustness and physical interpretability are both reduced. The backward interval yields a more stable feature space, a clearer predictor hierarchy, and a more direct connection to the underlying angular-momentum reservoir. This comparison provides additional justification for adopting $\Delta t_{-}$ as the primary temporal variable in the main analysis.

\begin{deluxetable}{lcclcclcc}[htbp]
\tablecaption{Stability scores for forward waiting-time classification.
\label{tab:forward_stability}}
\tablewidth{0pt}
\tablehead{
\colhead{Source} & \colhead{$N_{\rm g}$} & \colhead{Stability score} &
\colhead{Source} & \colhead{$N_{\rm g}$} & \colhead{Stability score} &
\colhead{Source} & \colhead{$N_{\rm g}$} & \colhead{Stability score}
}
\startdata
J1825$-$0935 & 13 & 0.368 & J1814$-$1744 &  7 & 0.369 & J0742$-$2822 &  8 & 0.392 \\
Crab         & 31 & 0.535 & J1341$-$6220 & 32 & 0.638 & J0631$+$1036 & 16 & 0.639 \\
J1801$-$2451 &  7 & 0.644 & J1826$-$1334 &  7 & 0.655 & J0205$+$6449 &  8 & 0.657 \\
J0537$-$6910 & 66 & 0.658 & J1841$-$0524 &  7 & 0.659 & J1413$-$6141 & 13 & 0.660 \\
J1420$-$6048 &  7 & 0.662 & J1801$-$2304 & 14 & 0.662 & J1023$-$5746 &  6 & 0.665 \\
J1740$-$3015 & 39 & 0.666 & J1731$-$4744 &  7 & 0.667 & J2229$+$6114 &  8 & 0.669 \\
J1737$-$3137 &  9 & 0.679 & J1048$-$5832 & 12 & 0.679 & Vela         & 25 & 0.723 \\
J1645$-$0317 &  6 & 0.793 & \nodata      & \nodata & \nodata & \nodata      & \nodata & \nodata \\
\enddata
\tablecomments{
Stability score measures the robustness of the forward waiting-time classification under bootstrap resampling. Lower and more dispersed values indicate reduced stability compared to the backward-interval case.
}
\end{deluxetable}

\clearpage
\bibliography{name}{}
\bibliographystyle{aasjournalv7}

\end{document}